\documentclass[pra,twocolumn,superscriptaddress,showpacs]{revtex4-1}
\usepackage{CJKutf8} 
\usepackage{amsmath,amssymb} 
\usepackage{bm}
\usepackage{hyperref}
\usepackage{subfigure}
\usepackage{graphicx}
\usepackage{color}

\begin{document}

\begin{CJK*}{UTF8}{} % Use default fonts from CJK (see below)

\title{Shell potentials for microgravity Bose-Einstein condensates}

\author{N.~Lundblad}
\email{nlundbla@bates.edu}
\affiliation{Department of Physics and Astronomy, Bates College, Lewiston, ME 04240, USA}

\author{R.A.~Carollo}
\affiliation{Department of Physics and Astronomy, Bates College, Lewiston, ME 04240, USA}

\author{C.~Lannert}
\affiliation{Department of Physics, Smith College, Northampton, Massachusetts 01063, USA}
\affiliation{Department of Physics, University of Massachusetts, Amherst, Massachusetts 01003-9300, USA}

\author{M.J.~Gold}
\affiliation{Department of Physics and Astronomy, Bates College, Lewiston, ME 04240, USA}

 \CJKfamily{gbsn} 
\author{X.~Jiang   (姜晓乐)}
\affiliation{Department of Physics and Astronomy, Bates College, Lewiston, ME 04240, USA}

\author{D.~Paseltiner}
\affiliation{Department of Physics and Astronomy, Bates College, Lewiston, ME 04240, USA}

\author{N.~Sergay}
\affiliation{Department of Physics and Astronomy, Bates College, Lewiston, ME 04240, USA}

\author{D.C.~Aveline}
\affiliation{Jet Propulsion Laboratory, California Institute of Technology, Pasadena, CA 91109, USA}

\date{\today}
\begin{abstract}

Extending the understanding of Bose-Einstein condensate (BEC) physics to new geometries and topologies has a long and varied history in ultracold atomic physics.    One such new geometry is that of a bubble, where a condensate would be confined to the surface of an ellipsoidal shell.    Study of this geometry would give insight into new collective modes, self-interference effects, topology-dependent vortex behavior, dimensionality crossovers from thick to thin shells, and the properties of condensates pushed into the ultradilute limit.   Here we discuss a proposal to implement a realistic experimental framework for generating shell-geometry BEC using radiofrequency dressing of magnetically-trapped samples.  Such a tantalizing state of matter is inaccessible terrestrially due to the distorting effect of gravity on experimentally-feasible shell potentials.  The debut of an orbital BEC machine (NASA Cold Atom Laboratory, aboard the International Space Station) has enabled the operation of quantum-gas experiments in a regime of perpetual freefall, and thus has permitted the planning of microgravity shell-geometry BEC experiments.     We discuss specific experimental configurations, applicable inhomogeneities and other experimental challenges, and outline potential experiments. 

\end{abstract}
\pacs{}
\maketitle

\end{CJK*}

\section{Introduction}

%
%* Nathan paper planning, old thoughts with Barry Garraay
%
%   OLD  Title: Shell trapping and enhancement of CAL experiment
%Working Title: Shell potentials for microgravity BECs
%
%DRAFT PAPER STRUCTURE:
%1. NL Introduction and Motivation 
%
%2. NL Dressed potentials, chips, and Trapping shells for CAL
%
% At the end we find inhomogeneities so on to the next section
%
%3. BG Enhancement of shell traps via microwave dressing
%
%4. NL/BG NLZ effects (subsection? - not sure)
%
% - NL will check calculation and see if we concur. Issue of would F=1
%   help in principle (but CAL is F=2 be aware).
%
% - F=1 vs. F=2 and sign of gF (Conclusion of NJP paper 2012)
% 
% - E.g. can NLZ non-linearity be reduced/changed to enhance trap
%   frequency. An alternate use of the microwave.

The study of quantum-degenerate ultracold atomic gases has historically been guided by explorations of geometry, dimensionality, topology, and interaction.   Whenever the parameter space of dimensionality and geometry has been expanded, interesting physics has typically been unveiled.   Studying Bose-Einstein condensates (BECs) in 2D has yielded insight into quasi-condensation and the BKT transition~\cite{Desbuquois:2012fh,Ha:2013jr,Hadzibabic:2008eo}, and in 1D insight into  fermionization and many-body systems out of equilibrium~\cite{Hofferberth:2007fk,Kinoshita:2006p765,Kinoshita:2004jp} .   Exploring toroidal condensates has driven progress in understanding persistent currents and uncovered links to cosmological inflation~\cite{Eckel:2018gv,Mathew:2015gg}, and double-well condensates have been used for many applications including matter-wave interferometry and spin squeezing\cite{Schumm:2005lr,Esteve:2008cl}.    A shell- or bubble-geometry BEC, while physically interesting due to its distinct topology, has not been physically realized due to the distorting influence of gravity on typical atom traps.    In this work we present modeling related to proposed experiments with bubble-geometry BECs aboard the NASA Cold Atom Laboratory (CAL), currently in operation aboard the International Space Station (ISS).  

An experimental path to creation of shell  potentials for BECs was proposed not long after the first creation of BEC itself, focusing on so-called adiabatic potentials created with radiofrequency-dressed magnetic traps~\cite{PhysRevLett.86.1195}, discussed further in Section \ref{rfdress}.   Alternate schemes for the study of shell BECs have focused on the specific study of the superfluid shells in optical-lattice Mott-insulator systems~\cite{barankov:063622}, or in the exotic environment of a neutron star~\cite{Pethick:2015ta}.     More recent theoretical work has focused on the collective modes of shell condensates, and the signatures of a  condensate transitioning  to a hollow shell from a conventional topology~\cite{Sun:2018de,Padavic:2017cv}.      Interesting effects are predicted to occur when a shell condensate is released into time-of-flight expansion; different regions of the shell BEC will interfere with each other, resulting in spatial matter-wave interference patterns that are quite sensitive to the shape of the shell potential and (via mean-field interactions) the number of atoms in the condensate~\cite{Lannert:2007kk}.      Recent work has also been done exploring the basic physics of BEC on the surface of a sphere\cite{Tononi:2019us,Bereta:2019th}.

Further, the motivation for the study of shell-shaped condensates stems from the drastic change in topology associated with expansion into a shell; vortex behavior (for example)  shows promise as an avenue of investigation, including the potential study  of vortex lattices in a curved background.  Vortices in a shell-shaped condensate will behave in a qualitatively different manner than those in a flat condensate (such as a disk) because of the curvature of the shell surface and because of the topology of the shell as an unbounded simply-connected surface.   Previous theoretical work has predicted that a single pair of vortices on the surface of a sphere will repel and therefore arrange themselves at polar-opposite points~\cite{Milagre:2007jt}. Vortices in the shell-condensate system can be induced through rotations or, if the shell is thin enough, they will be spontaneously produced near the thermal transition to a non-condensed gas~\cite{Hadzibabic:2006lr}.   The effect of curvature on vortices in a thin condensate is a richer area for exploration; for example, defects (such as vortices) on a curved surface experience a force due to the local curvature~\cite{Turner:2010kl,Vitelli:2004gv}.   

%%%
\section{Radiofrequency dressing}\label{rfdress}

The interaction of radiofrequency (rf) or microwave radiation with a set of Zeeman-split hyperfine manifolds is a well-studied system that is often characterized in terms of dressed states, which in the case of inhomogeneous magnetic fields such as found in magnetic traps result in so-called adiabatic potentials~\cite{Garraway:2016hu,perrin2017trapping}.    Figure 1 illustrates the general idea of radiofrequency dressing; lower-lying adiabatic potentials are associated with the ``rf knife" techniques of evaporative cooling in magnetic traps, while the higher-lying adiabatic potentials can be understood as double-wells in 1D, ring potentials in 2D, and shell potentials in 3D, as first proposed by Zobay and Garraway~\cite{zobay:023605,PhysRevLett.86.1195,Zobay:2000tk}.   Experimentally, ultracold gases in rf-dressed shell potentials were first generated with the key observation that gravitational sag caused the shell-trapped samples to localize near the bottom of the shell potential~\cite{white:023616,perrinbec}; indeed, this localization could be considered a feature due to the possibilities of applying it to studies of effectively 2D quantum gases~\cite{Merloti:2013ft}.  

To calculate the dressed potentials associated with a single driving frequency $\omega$ of coupling strength $\Omega$, we operate in the usual rotating frame~\cite{perrin2017trapping} and take the rotating-wave approximation resulting in the Hamiltonian 

\begin{widetext}

\begin{equation}\label{ham}
{\mathcal H}=\left( \begin{matrix}
2\omega & \Omega/2 & 0 & 0 & 0 \\ 
\Omega/2 & \omega & \frac{\sqrt{3}}{2}\Omega/2  & 0 & 0 \\
0 &\frac{\sqrt{3}}{2}\Omega/2& 0 &\frac{\sqrt{3}}{2}\Omega/2& 0\\
0 & 0 & \frac{\sqrt{3}}{2}\Omega/2 & -\omega &\Omega/2 \\
0 & 0 & 0 & \Omega/2 & -2\omega \\
\end{matrix}\right) + {\mathcal H_{\rm Zeeman}({\bf r})}  
\end{equation}

\end{widetext}
where $ {\mathcal H_{\rm Zeeman}({\bf r})} $ is diagonal and represents the (exact) Zeeman shifts of the  states in use, which for the purposes of this work are the $^{87}$Rb upper hyperfine ground state denoted by  $|F=2, m_F \rangle$, with $m_F$ taking values from -2 to 2.    Modeling of terrestrial experiments would require the addition of an $m g z$ term to ${\mathcal H}$.    The spatially-varying eigenvalues of the Hamiltonian in Eq.~1 represent the adiabatic potentials  and the eigenvectors represent the spatially-varying decomposition of the dressed state in the lab-spin basis.

%%%%%%%%%%%%%%%%%%%%%%%%
\begin{figure}[b!]
\centering
  \includegraphics[width=\columnwidth]{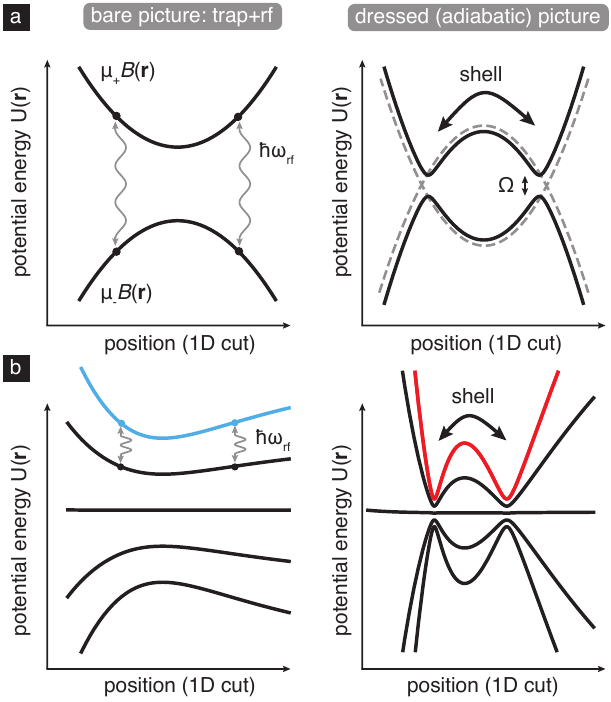}\caption{
 Schematic representation of rf dressing and bubble-potential creation.   {\bf (a)} A  two-level system in a magnetic trap undergoing rf dressing; at left, the bare picture of a trapped state and an antitrapped state with an rf signal resonant at a particular point in space.    {\bf (b)}  A realistic five-level system ($F=2$) in an atom-chip style magnetic trap as discussed later.     The initially trapped state (blue) can be adiabatically converted to the dressed state (red), depicted at right.   In both cases energy levels are split by Zeeman shifts in the bare picture and by coupling strength $\Omega$ in the dressed picture.   The shell potential can be visualized by considering these curves as slices through a 3D potential; thus, trapped atoms will reside at the `notches' of the curves at right. }     
\label{dressing}
\end{figure}
%%%%%%%%%%%%%%%%%%%%%%%%

The detuning of the rf field acts to control the mean radius of the bubble potential, and the coupling strength $\Omega$ (which could have some weak spatial dependence) serves the twofold purpose of controlling the curvature of the local bubble minimum but also ensuring (through sufficiently large magnitude) stability against Landau-Zener-type nonadiabatic losses in this dressed-state picture.     These losses have been explored in the context of magnetic traps~\cite{Burrows:2017bw} and also connected to the stability of condensates in radiofrequency-dressed spin-dependent optical lattices~\cite{Lundblad:2014cc,Lundblad:2008kb}.

%%%
\section{Cold Atom Laboratory (CAL) Apparatus}

The rf-dressing process resulting in shell-like BEC could be performed in any ultracold atomic physics experimental framework featuring magnetic trapping and elimination of gravitational perturbation, and thus could be implemented in drop-tower~\cite{Muntinga:2013ge,VanZoest:2010p3350}, ballistic aircraft~\cite{Barrett:2016ko}, or the most recently developed sounding-rocket~\cite{Becker:2018db} configurations (the latter representing the first BEC experiment in space).    Our investigation has focused on planned experiments aboard the NASA Cold Atom Laboratory (CAL).   CAL was developed by the Jet Propulsion Laboratory (JPL) beginning in 2013 and is currently in operation aboard the International Space Station (ISS) after a 2018 delivery via a Cygnus spacecraft launched from NASA's Wallops Flight Facility.    CAL is an atom-chip-based BEC machine equipped with a variety of experimental degrees of freedom permitting operation with multiple experimental PIs with a diversity of experimental frameworks, including Efimov physics~\cite{Mossman:2016vpa}, adiabatic expansion and delta-kicked cooling to pK temperatures~\cite{Sackett:2017gp,Myrskog:2000il}, novel atom lasers~\cite{Meister:2019bv}, and ongoing development of atom-interferometer capabilities.

%%%%%%%%%%%%%%%%%%%%%%%%
\begin{figure}[t!]
\centering
  \includegraphics[width=\columnwidth]{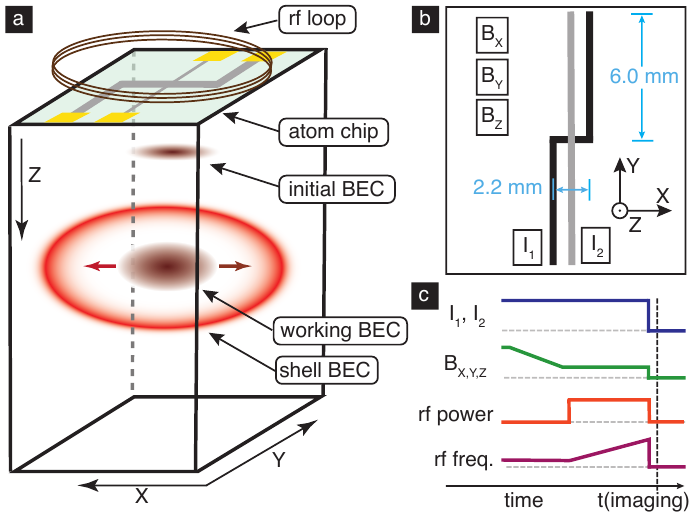}\caption{Design of atom-chip apparatus and basis for modeling.   {\bf (a)} the CAL science chamber, depicting essential components, and the planned state of the atomic cloud at various points in the experimental cycle; initial tight trap provided by standard CAL procedures, decompressed `working'  trap with lower aspect ratio and increased distance from chip surface, and inflated shell BEC, adiabatically converted from the working trap.   {\bf (b)} schematic of the modeled aspects of the atom chip, comprising two chip currents and three bias fields.  {\bf (c)} planned experimental sequence for generating ultracold gases in a shell potential; from top: chip currents, bias fields providing initial decompression; rf dressing power (turned on far from resonance); rf dressing frequency, ramped upwards to inflate the shell potential.     Absorption imaging occurs immediately after rapid switchoff of chip currents, bias fields, and rf.  
}    
\label{hardware}
\end{figure}
%%%%%%%%%%%%%%%%%%%%%%%%

%%%%%%%%%%%%%%%%%%%%%%%%
\begin{figure*}[t!]
\centering
  \includegraphics[width=\textwidth]{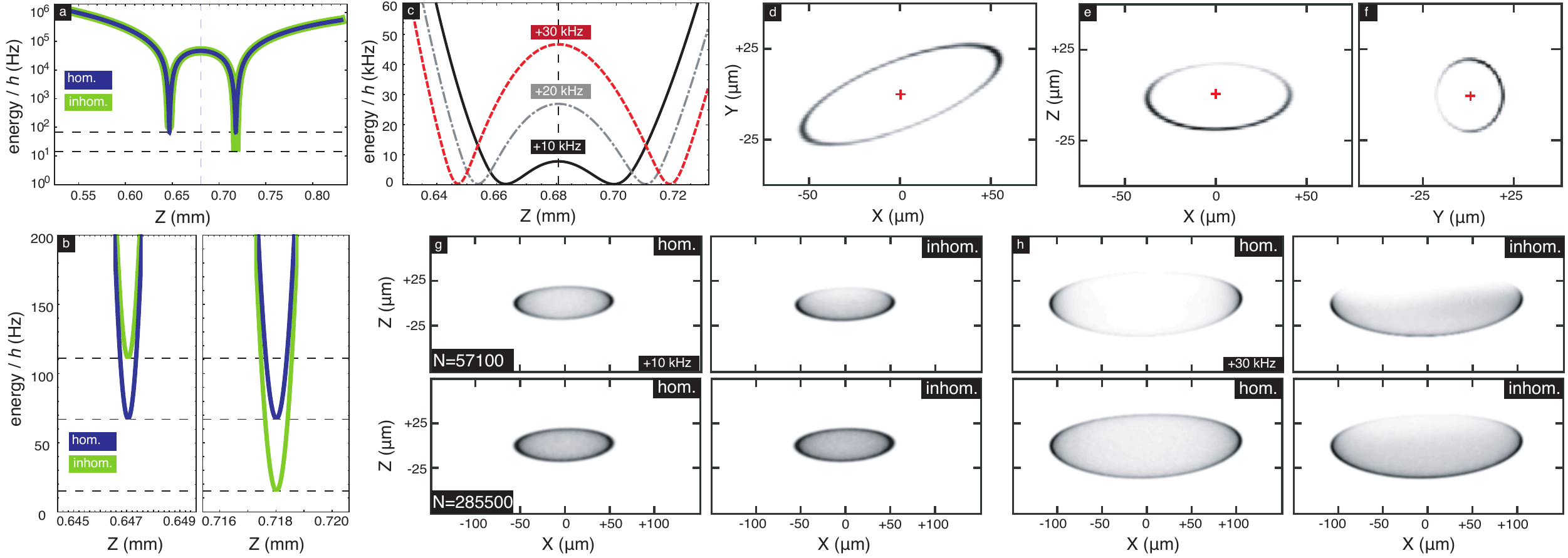}\caption{Realistic modeling of Bose-Einstein condensates in a typical CAL atom-chip potential.   {\bf (a)} A slice of potential energy $U(x=0,y=0,z)$ associated with a dressed trap configuration $I_1=$3.2 A, $I_2=-.4$ A, $B_x=$2.07 G, $B_y=$7.11 G, $B_z=$0.37 G.    The trapping potential is associated with the uppermost adiabatic potential of the $F=2$ ground-state manifold in $^{87}$Rb.    The curve labeled ``inhom." takes into account the inhomogeneity of the rf coupling strength, resulting in a smaller avoided-crossing gap further from the atom chip, and thus an effective trap tilt in this configuration of $\sim h\times100$ Hz.    {\bf (b)} Fine detail of the local minima in (a).   This residual tilt is equivalent to $\sim.001g$.  {\bf (c)} Variation of the shell potential with rf detuning $\Delta$; $\Delta=0$ associated with magnetic resonance at trap bottom.    Note increasing shell radius with detuning.     {\bf (d--f)}.   Modeled slices in the principal experimental planes of atomic density $n({\bf r})$ associated with $N=57100$, detuning $+10$ kHz, with inhomogeneities included (i.e.~associated with the column density represented at upper right in (g)).     Note impact of inhomogeneity partially suppressing atomic density at $+z$ and $-y$.       {\bf (g--h)} A variety of modeled column-density images, associated with the absorption imaging direction ($y$) on the CAL instrument.     Low detuning and small shell radius are represented in (g), with higher detuning and correspondingly larger shell radius in (h).      A factor of 5 change in atom number $N$ is associated with movng from the top to the bottom row; this change results in more uniform filling of the potentials.      All images in (d--h) are scaled to peak density; for all data in this Figure the rf coupling strength is set to  $\Omega/2\pi=$  5 kHz.  
}  
\label{pots}
\end{figure*}
%%%%%%%%%%%%%%%%%%%%%%%%

General capabilities of the instrument (see accompanying illustrations in Fig.~\ref{hardware}) include providing  $^{87}$Rb BECs with $N > 10^4 $ in an initial high-aspect ratio trap configuration with approximate trap frequencies $\{\omega_x,\omega_y,\omega_z\}=2\pi\times\{200,1000,1000\}$ Hz, where $z$ is the direction perpendicular to the atom chip and $x$ is the direction associated with the bias magnetic field at trap bottom.    Condensates are obtained via rf evaporation of a sample held in the magnetic trap formed by a combination of currents flowing through the atom-chip wires and three quasi-uniform external bias fields.  Details of system development and ground test status can be found in Ref.~\cite{Elliott:2018dk}.   Specific design input was sought from prospective users; for example, significant guidance regarding the rf system design of dressed-atom experiments can be found in the literature, specifically focusing on the need for DDS signal sources and very fine-grained frequency ramps during the dressing process in order to avoid excess heating~\cite{Morizot:2008bq}.   A key capability to begin dressed-atom experiments with CAL is the generation of traps of lower density and aspect ratio; hence, a trap expansion protocol that does not incur unwanted center-of-mass motion is desired.   Such paths have been developed in the context of shortcuts to adiabiaticity with drop-tower missions~\cite{Corgier:2018iu} and in planning for CAL; we have developed expansion ramps roughly in the form of a hyperbolic tangent, following the formalism of Ref.~\cite{Sackett:2017gp}.  

The general procedure for forming a shell condensate in a machine such as CAL would be as follows, as parametrized in Fig.~2(c).   First, the condensate would be prepared in a given initial starting condition (the ``bare trap", in the internal state $|F=2,m_F=2\rangle$), at which point the rf dressing signal would be switched on with the detuning $\Delta = \omega-\omega_0$ negative and large compared with the Rabi frequency $\Omega$, where $\omega_0$ is associated with magnetic resonance at trap bottom.    Secondly, the rf frequency would be ramped upwards, forcing the condensate in the uppermost adiabatic potential into a shell geometry.    The timescale of this ramp would be enforced by mechanical adiabaticity of the BEC deformation and technical limits on the graining of the rf signal; timescales associated with motion perpendicular to the local shell surface are easily satisfied, but adiabaticity with respect to motion around the shell remains an open question.   
%Typical dressing ramps found in terrestrial experiments anticipated to be appropriate for CAL are of order $d\omega/dt =$ 1--10 kHz/ms.
    Coupling strengths $\Omega/2\pi \sim$ 10 kHz are appropriate for these scenarios, chosen in the context of the suppression of Landau-Zener losses~\cite{Merloti:2013ft}; this rf amplitude is well within the documented capability of the CAL instrument.
%%%

\section{Ground state \& inhomogeneities}\label{gs}

Following the formalism summarized by the Hamiltonian in Eq.~\ref{ham} we calculate adiabatic potentials for several different cases of interest.   In particular, it is useful to investigate the effects of rf detuning and atom number on the planned experiments, and explore the consequences of various inhomogeneities associated with the experiment.   The most dominant inhomogeneity associated with such experiments on Earth is gravitational potential energy $m g z$ (absent in Eq.~1), which for $^{87}$Rb corresponds to a tilt of $h\times$2.14 kHz/$\mu$m  (or $k_B\times$103 nK/$\mu$m).  Taken into account across a typical condensate profile, this dwarfs the ability of BEC interaction energy to ``fill up" a gravitationally tilted shell.  In freefall this effect is absent and we are left with several confounding factors orders of magnitude smaller, framed as follows: inhomogeneity A, associated with the ellipsoidal aspect ratio of the shell potential, inhomogeneity B, associated with the wandering of the local magnetic field direction across the trapped atomic cloud (impacting dressing via departure from orthogonality with the dressing field), and inhomogeneity C, associated with the difference in $\Omega$ across the sample.    

While slices of condensate density along principal directions are useful modeling checks, experimental data will come in the form of column density along a particular direction.     Fig.~\ref{pots} shows calculated condensate density slices (\ref{pots}d--f) and column densities (\ref{pots}g--h) for planned magnetic field configurations.  The example trap chosen has identical atom-chip currents to the ``tight trap" where the CAL BEC first forms, but has had external bias field reduced to 0.2$\times$ their initial value, resulting in trap frequencies of approximately (30, 100, 100) Hz as suggested by our model and by initial calibration experiments aboard CAL.    Also shown are examples of column densities taken without accounting for the inhomogeneities A, B, C as defined above, to illustrate their impact.    The coupling-related inhomogeneity C pulls the condensate toward  $+z$ (away from the chip); inhomogeneity B pulls toward $+y$, and inhomogeneity A results in pooling of atoms at the tips of the trap ellipsoid.

 Condensate densities are calculated using an imaginary-time propagation-based Gross-Pitaevskii solver~\cite{Chiofalo:2000ky,Cerimele:2000hs} using the uppermost dressed-stated potential $U({\bf r})$ as input, along with illustrative condensate numbers $N$ chosen to be different by a factor of 5 and consistent with CAL specifications.  These calculations confirm typical intuition, that the Gross-Pitaevskii nonlinearity driven by repulsive atom-atom interaction (i.e.~the chemical potential $\mu$) serves to some degree to conceal nonuniformities  that are of order $\mu$.     However, this benefit is limited; the ground-state energies associated with the the scenarios in Fig.~3(g-h) range from $h\times$ 100--200 Hz (or $k_B\times$ 5--10 nK), an order of magnitude smaller than the ground-state energy of the original condensates.    In general, atom number $N$ is not large enough in the CAL scenario to drive a shell-trapped condensate into the interaction-dominated Thomas-Fermi regime.   Nevertheless, the Gross-Pitaevskii ground states show that a shell-trapped BEC (of size $\sim 50~\mu$m) is within the capabilities of the CAL system to observe, with the caveat that complete density coverage around the surface of the shell will be strongly sensitive on the atom number made available.     The effects of terrestrial gravitational tilt is absent in these plots, given the planned microgravity environment; were it present, the modeled clouds would be very strongly pinned to one end of the trap as in the terrestrial experiments~\cite{Harte:2018jm,white:023616,perrinbec}.    The significant inhomogeneity C (that of the rf coupling) can be reduced to some degree by moving to lower absolute coupling strength, given that the tilt is proportional to $\Omega$; this would be at the eventual cost of reduced dressed-state lifetime due to Landau-Zener nonadiabaticity.  For future experiments, it also could be mitigated through experimental design (e.g.,~rf loop radius and placement). 

%Relating these modeled condensate column densities to observations will require clear understanding of the NASA CAL imaging system and how it interacts with samples such as the shell traps discussed here.   Additional calibration of the model potential with respect to background fields will also be required, as will precise measurement of the coupling strength $\Omega$.  

%%%
\section{Future prospects}

Following five years of development, NASA CAL was recently commissioned aboard ISS after 2018 launch.  It has undergone testing and is in active user-facility mode with several PI groups, including the authors. Initial work will focus on calibration of the various models used to predict trap fields, trap frequencies, and other properties of the atom-chip system, followed by exploration of residual motion in given trap configurations where the magnetic trap has been expanded and translated away from the chip surface.   Assuming sufficiently stable BEC production, stable trap position, and repeatable magnetic resonance observations, rf dressing of the CAL atom-chip trap is within reach. 

Beyond confirmation of shell structure the microgravity BECs, discovery-oriented user time will focus on elucidation of the adiabaticity requirements of shell creation, possible exploration of collective-mode dynamics, and studies of the lifetime of BEC shell structures.   We also anticipate observation and characterization of the inhomogeneities predicted and discussed in Section.~\ref{gs}, and exploration of the behavior of non-condensed thermal atoms in the dressed potential, depending on what condensate fractions are available on orbit.   To frame future design considerations, we note that shell thickness might potentially be tunable through use of the nonlinear Zeeman shift \cite{SinucoLeon:2012bl}, and that the inhomogeneity associated with the rf loop could potentially be compensated through application of a similarly inhomogeneous microwave dressing field, i.e.~a compensatory ac Zeeman shift~\cite{SinucoLeon:2019ue, Garraway:dikHLmv0}.

CAL is currently scheduled to remain in operation until late 2019, whereupon a major hardware replacement is scheduled to occur, after which the facility should return to user-facility operations for additional time.  A second-generation orbital microgravity atom-chip ultracold atomic physics facility,~BECCAL, is currently under development of in Germany as a joint DLR/NASA venture~\cite{Becker:2018vc}.  This successor machine should share CAL's capabilities for generation of rf-dressed systems, enabling a second-generation exploration of shell-BEC physics.

\vspace{.1in} 

\begin{acknowledgments}

We thank Ethan Elliott, Barry Garraway, Jeffrey Oishi,  German Sinuco-Le\'{o}n, Smitha Vishveshwara, and Karmela Padavi\'{c} for useful discussion.   This work was supported by Jet Propulsion Laboratory (JPL) Research Support Agreement No.~1502172 under a contract with the National Aeronautics and Space Administration (NASA), Division of Space Life and Physical Sciences Research and Applications  (SLPSRA).    

\end{acknowledgments}
\bibliography{}

%\bibliographystyle{apsrevNOURL}
%\bibliography{/Users/nlundbla/Documents/LaTeX/nl-master-4}

\end{document}